\documentclass[prd,showpacs,preprintnumbers,amsmath,amssymb,superscriptaddress,floatfix,nofootinbib]{revtex4}
\usepackage{graphicx}
\usepackage{epstopdf}
\usepackage{amsmath}
\usepackage{amsfonts}
\usepackage{amssymb}
\usepackage{bm}
\usepackage{xcolor}

\begin{document}

\title{Searching for a $D \bar D$ bound state with the $\psi(3770) \to \gamma D^0 \bar D^0$ decay}
\author{Lianrong Dai}
\email[]{dailianrong68@126.com}
\affiliation{Department of Physics, Liaoning Normal University, Dalian 116029, China}

\author{Genaro Toledo }
\email[]{toledo@fisica.unam.mx}
\affiliation{Instituto de F\'{\i}sica, Universidad Nacional Aut\'onoma de M\'exico, AP 20-364, Ciudad de M\'exico 01000, M\'exico.}
\affiliation{Departamento de F\'{\i}sica Te\'orica and IFIC,
Centro Mixto Universidad de Valencia-CSIC Institutos de Investigaci\'on de Paterna, Aptdo.22085, 46071 Valencia, Spain}

\author{Eulogio Oset}
\email[]{oset@ific.uv.es}
\affiliation{Departamento de F\'{\i}sica Te\'orica and IFIC,
Centro Mixto Universidad de Valencia-CSIC Institutos de Investigaci\'on de Paterna, Aptdo.22085, 46071 Valencia, Spain}
\date{\today}

\begin{abstract}
\noindent	
  We perform a calculation of the mass distribution in the $\psi(3770) \to \gamma D \bar D$ decay, studying both the  $D^+  D^- $ and $D^0 \bar D^0 $ decays. The electromagnetic interaction is such that the tree level amplitude is null for the neutral channel, which leaves the $\psi(3770) \to  \gamma D^0 \bar D^0$ proceeding through a loop involving the $D^+ D^- \to D^0 \bar D^0$ scattering amplitude. We use the results for this amplitude of a theoretical model that predicts a $D \bar D$ bound state and find a  $D^0 \bar D^0 $ mass distribution in the decay drastically different than phase space. The rates obtained are relatively large and the experiment is easily feasible in the present BESIII facility. The performance of this experiment could provide an answer to the issue of this much searched for state, which is the analogue of the $f_0(980)$ resonance.
\end{abstract}



\maketitle
\section{Introduction}
   Molecular states made from mesons or mesons and baryons have become some of the important objects in the present plethora of hadronic states. Reviews on this topic can be found in \cite{ramosoller,review} and more recently in \cite{olsen,karliner,huaxing,fengkun}. One of the interesting predicted states is a bound state of $D \bar D$ \cite{dani,pavon,hidalgo} for which there is no clear experimental evidence so far. The state is analogous to the $K \bar K$ bound state which was claimed in \cite{isgur} to be a good representation of the $f_0(980)$ state. This early claim found support later with the results of the chiral unitary approach for the meson-meson interaction \cite{npa,kaiser,markushin,juanenri}, where the meson meson interaction in coupled channels was studied, and other states, as the $f_0(500)$, $K^*_0(700), a_0(980)$, were also found dynamically generated from that interaction.

   There has been some search for this state and in \cite{daniee} it was shown that an accumulation of strength close to the $D \bar D$ threshold in the  $ e^+e^-\to J/\psi D \bar D$ reaction \cite{pakhlov} found a natural explanation in terms of  the predicted bound state of \cite{dani} with a mass of 3730 MeV, with large uncertainties. Hopes where raised that an update of the experiment in \cite{chilikin} would constrain the predictions, but it was shown in \cite{wangliang} that this is not the case, and there is a large ambiguity in the conclusions.

    In view of this, there have been some works proposing new reactions that would give evidence for this elusive state. In \cite{danizou} the radiative decay of the $\psi(3770)$ resonance into $\gamma$ and the $D \bar D$ bound state was proposed and the feasibility of the reaction with present production rates of the $\psi(3770)$ was assessed. There is of course the problem of which should be the ideal channel to observe the bound state. In \cite{xiaothree} three different reactions were suggested to detect that state. In \cite{daixie} the $B^0 \rightarrow D^0 \bar{D}^0 K^0$ , $B^+ \rightarrow D^0 \bar{D}^0 K^+$ reactions were suggested to find evidence for the state, looking into the mass distribution of $D \bar D$ production close to threshold. The analysis found a good agreement with experiment for the $B^+ \rightarrow D^0 \bar{D}^0 K^+$ reaction, but it was shown there that the $B^0 \rightarrow D^0 \bar{D}^0 K^0$ reaction was better suited to search for the $D \bar D$ state, because there is no tree level contribution in this later reaction, and hence the amplitude for that mechanism is proportional to the $D \bar D$ amplitude which contains the pole of the   $D \bar D$ bound state. The mass distribution then was rather different from that of phase space, and its precise  measurement close to threshold should give and answer to the question.

  In the present work we wish to combine the lessons drawn from the works of \cite{danizou} and \cite{daixie} and study the $D \bar D $ mass distribution close to threshold from the $\psi(3770) \to \gamma D \bar D$ decay.  Anticipating the results, we will find an interesting situation in which the $D^0 \bar D^0 $ production does not proceed at tree level, while  $D^+ D^- $ has contribution from tree level. As a consequence, the $D^0 \bar D^0 $ production is directly influenced by the $D \bar D$ pole below threshold and exhibits a behavior close to threshold very different from phase space. For the $D^+ D^- $ production the tree level part is very important and the behavior is quite different, and in addition it shows the infrared divergence behavior when the photon energy goes to zero, which, again, is not the case for $D^0 \bar D^0 $ production. The reaction is, thus, suited for investigation of the $D \bar D$ bound state and the present rates of $\psi(3770)$ production make the experimental investigation feasible.

\section{Formalism}
The state $\psi(3770)$ decays into $D\bar D$ \cite{pdg} with a width $\Gamma=27.2$ MeV,  52\% of which goes to $D^0 \bar D^0$ and 41\% to $D^+D^-$. Its shape in $e^+e^-$ production and the decay width have been object of intense study \cite{bes1,bes2,babar,coito, qzhao1,qzhao2,qixin}.
In \cite{qixin} the $c\bar c$ component is allowed to get hadronized into meson-meson components, and the strength  of the hadronization is fitted to the $\psi(3770)$ lineshape. One of the conclusions in \cite{qixin} is that from the experimental data one can induce that the $\psi(3770)$ is largely a $c\bar c$ state and the weight of the meson-meson components is only of the order of 15\%. The $\psi(3770)$ state bears some similarity to the $\phi(1020)$, which is assumed to be a $s\bar s$ state and decays into $K\bar K$. The decay mode $\psi(3770) \to D \bar D \gamma$ necessarily has much resemblance to the
$\phi \to \gamma K \bar K, \  \gamma \pi^0 \pi^0$ decays, which have also been a subject of much study \cite{bramon,kim,colan,achasov,pancheri,lucio,marco,palomar}.
As in the $\phi \to  \gamma \pi^0 \pi^0$ reaction, which  does not proceed via tree level, we shall also see that $\psi(3770) \to \gamma D^0 \bar D^0 $ does not get contribution from tree level and both processes proceed via  a similar loop mechanism.

\subsection {Tree level for $\psi(3770) \to \gamma D^+ D^- $}
\begin{figure}[h!]
\begin{center}
\includegraphics[scale=0.8]{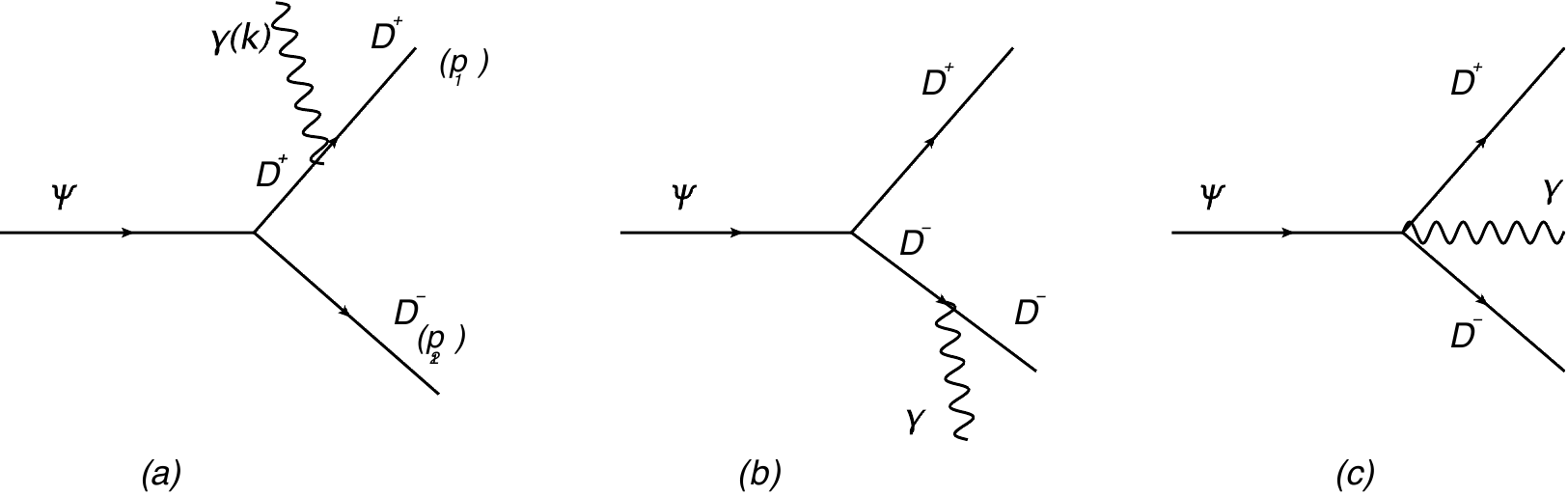}
\end{center}
\vspace{-0.7cm}
\caption{Mechanisms for $\psi \to \gamma D^+ D^- $: (a), (b), $D$ pole mechanisms; (c) contact term demanded by gauge invariance. In parenthesis the momenta of the particles.}
\label{Fig:1}
\end{figure}

The tree level mechanism in $\psi(3770) \to \gamma D^+ D^- $ is shown diagrammatically in Fig. \ref{Fig:1}.

The $\psi \to D^+ D^- $ elementary vertex is given by
\begin{equation}\label{eq:vertex}
-it_{\psi D^+D^-} = -i g_\psi (p_{D^+}-p_{D^-})_\mu \epsilon^\mu(\psi)  \,.
\end{equation}

The $\psi \to D^+ D^- $ decay width is given by
\begin{equation}\label{eq:width}
\Gamma_\psi=\frac{1}{8\pi}\frac{1}{M_\psi^2}|{\bm q}| \overline{\sum} \sum\vert t \vert^2 \,,
\end{equation}
where the sum and average of $|t|^2$ calculated from Eq. (\ref{eq:vertex})
gives
\begin{equation}\label{eq:sum}
\overline{\sum}\sum \vert t \vert^2=\frac{4}{3} \, g_\psi^2 \,{\bm q}^2 \,,
\end{equation}
and ${\bm q}$ is the $D^+$ momentum in the $\psi$ decay at rest. Adjusting to the experimental $D^+D^-$ decay width we find
\begin{equation}\label{eq:coupling}
g_\psi=13.7
\end{equation}
Considering also the $\gamma D^+D^-$ coupling $D^+(p_{D^+}) \gamma \to D^+(p'_{D^+})$
\begin{equation}\label{eq:VDDgamma}
it_{\gamma D^+D^-}=-i\,e\,(p_{D^+}+p'_{D^+})_\mu \,\epsilon^\mu(\gamma)  \,,
\end{equation}
with $e$ the electron charge, $e^2/4\pi=\alpha=1/137$,
the  $\psi \to \gamma D^+ D^- $ amplitude of the diagram of Fig. \ref{Fig:1}
is given by
\begin{eqnarray}\label{eq:treeamplitude}
t_a+t_b+t_c =
-2\,e\,g_\psi\,\epsilon^\mu(\psi) \epsilon^\nu(\gamma)
\left( g_{\mu\nu}+{p_2}_\mu {p_1}_\nu \frac{1}{p_1\cdot k+i\epsilon}
+{p_1}_\mu {p_2}_\nu \frac{1}{p_2\cdot k+i\epsilon}
\right) \,,
\end{eqnarray}
where the term $g_{\mu\nu}$ corresponds to the diagram of Fig. \ref{Fig:1}(c)
and is introduced to respect gauge invariance.
The photon has zero coupling to $D^0 D^0$ and hence there is no tree level for $\psi \to \gamma D^0 \bar D^0$.

\subsection{Loop mechanism}
\begin{figure}[h!]
\begin{center}
\includegraphics[scale=0.95]{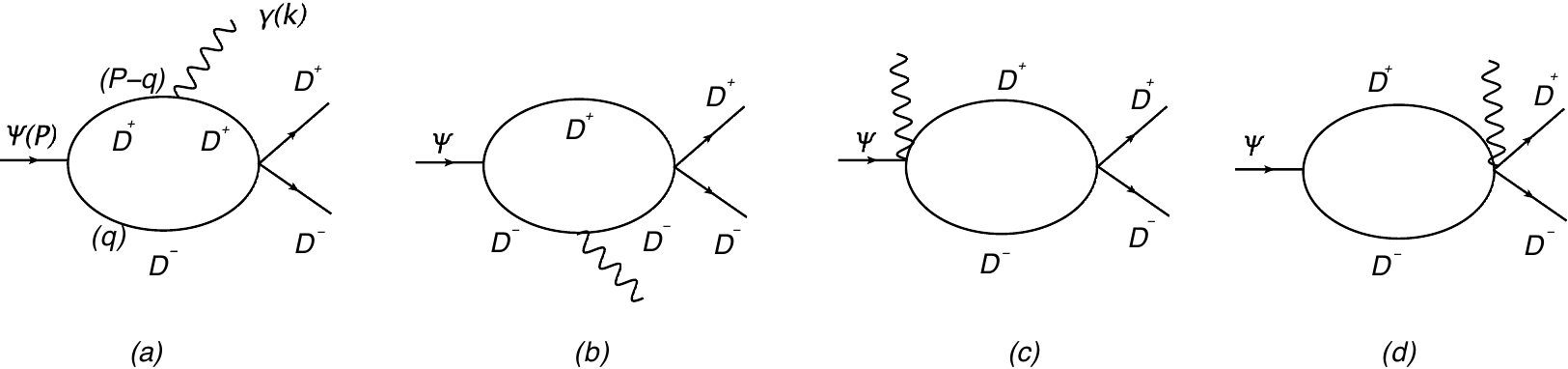}
\end{center}
\vspace{-0.7cm}
\caption{Loop mechanisms for $\psi \to \gamma D \bar D $ production. In parenthesis the momenta of the particles.}
\label{Fig:2}
\end{figure}

There is, however, a loop mechanism that allows the $\psi \to \gamma D^0 \bar D^0$ decay which is depicted in Fig. \ref{Fig:2}.

This follows exactly the same trend as in \cite{bramon,kim,colan,achasov,pancheri,lucio,marco,palomar} for $\phi \to \gamma \pi^0 \pi^0$, where the intermediate state is $K^ + K^-$ and the final $D\bar D$ are replaced by $\pi^0 \pi^0$. The diagram of Fig. \ref{Fig:2}(d) is also demanded by gauge invariance of the loops. Gauge invariance plays an important role in this process and thanks to it there is an efficient computational scheme which requires only the evaluation of diagrams  (a) and (b) of Fig. \ref{Fig:2}, which give the same contribution, and shows that the result of the loop integral is convergent \cite{pancheri,pestiu,close}. The derivation goes as follows: The full amplitude for the diagrams of Fig. \ref{Fig:2} has the structure
\begin{equation}\label{eq:loopamp}
t_L= \epsilon_\mu(\psi)\epsilon_\nu(\gamma)\,T^{\mu\nu}
\end{equation}
and $T^{\mu\nu}$  must be a tensor that can be written in terms of the two independent momenta $P$ and $k$, the momentum of the $\psi$ and $\gamma$ respectively. The most general form for
$T^{\mu\nu}$ is given by
\begin{equation}\label{eq:Tmunu}
T^{\mu\nu}=a\,g^{\mu\nu}+ b\,P^\mu P^\nu+  c\,P^\mu k^\nu+ d\,k^\mu P^\nu +e\,k^\mu k^\nu \,.
\end{equation}
Gauge invariance, substituting $\epsilon_\nu(\gamma)$ by $k_\nu$ and demanding
\begin{equation}\label{eq:gaugeinv}
T^{\mu\nu} k_\nu=0
\end{equation}
leads to
\begin{equation}\label{eq:TmunuGI}
a\,k^{\mu}+ b\,P^\mu (P\cdot k)+ d\,k^\mu (P\cdot k)=0
\end{equation}
which implies two independent equations
\begin{eqnarray}\label{eq:conditions}
a+d\,(P\cdot k)=0\\
b=0.
\end{eqnarray}
The $b$ term does not contribute because of the Lorentz condition $\epsilon_\mu(\psi) P^\mu=0$ and the $c$ and $e$ terms do not contribute because of the Lorentz condition on the photon $\epsilon_\nu(\gamma) k^\nu=0$. Hence, only the $a$ and $d$ terms of Eq. (\ref{eq:Tmunu}) contribute to the amplitude and it is enough to calculate only the $a$ or $d$ coefficient. It is easy to see that only the diagrams (a), (b) of Fig. \ref{Fig:2} contribute to the $d$ coefficient and since two external momenta $P^\mu k^\nu$ are factorized out of the integral, for dimensional reasons this means two powers of $q$ less in the integral, which renders it convergent. In addition, if we work at the end, as we do, in the Coulomb gauge, $\epsilon^0(\gamma)=0$, ${\bm \epsilon}(\gamma)\cdot {\bm k}=0$, then the term $dP^i k^j \epsilon_j(\psi)\epsilon_i(\gamma)=0$ in the $\psi$ rest frame, and the whole amplitude is given by
\begin{equation}\label{eq:finalamp}
t=a \epsilon_\mu(\psi)\epsilon^\mu(\gamma) \,;\qquad  \ a =-d\,(P \cdot k) \,
\end{equation}
It is customary to perform the integration of the loop integral using Feynman parametrization, but here we must divert from this formalism because the $D\bar D \to D\bar D$ scattering matrix regularized with a cut off, $q_{max}$, transfers a structure $\Theta(q_{max}-|{\bm q}|)\Theta(q_{max}-|{\bm p}|)$ to the $T$ matrix \cite{danijuan} and we must implement a cut off in the loop integral. On the other hand, we can benefit from the fact that the $D$ mesons are heavy particles, they are close to on-shell in the loops and we can just keep the positive energy part of their propagators
\begin{eqnarray}
D({q}) \to \frac{1}{q^2-m_D+i\epsilon} &\equiv&
\frac{1}{2\omega({\bm q})}
\left(
\frac{1}{q^0-\omega({\bm q})+i\epsilon}
-\frac{1}{q^0+\omega({\bm q})-i\epsilon}
\right)\label{eq:propagator}\\
&\to&
\frac{1}{2\omega({\bm q})}
\frac{1}{q^0-\omega({\bm q})+i\epsilon}
\end{eqnarray}
with $\omega({\bm q})=\sqrt{{\bm q}^2+m_D^2}$.\\

The contribution of the two diagrams of Fig. \ref{Fig:2}(a), \ref{Fig:2}(b) with $D^0\bar D^0$ in the final state is given by
\begin{eqnarray}\label{eq:amp1}
-it_L&=&2\int
\frac{d^4q}{(2\pi)^4}
(-i)g_\psi (P-q-q)_\mu  \epsilon^\mu(\psi)
(-ie)(P-q+P-q-k)_\nu  \epsilon^\nu(\gamma) (-i)t_{D^+D^- \to D^0\bar D^0}\nonumber\\
&&\times \frac{i}{(P-q)^2-m_D^2+i\epsilon}
\frac{i}{q^2-m_D^2+i\epsilon}
\frac{i}{(P-q-k)^2-m_D^2+i\epsilon},
\end{eqnarray}

\begin{eqnarray}\label{eq:amptot4}
t_L&=&-2e g_\psi \epsilon^\mu(\psi) \epsilon^\nu(\gamma) i \int \frac{d^4q}{(2\pi)^4}  2q_\mu (2P-2q)_\nu t_{D^+D^- \to D^0\bar D^0} \nonumber\\
&&\times \frac{1}{(P-q)^2-m_D^2+i\epsilon}\frac{1}{q^2-m_D^2+i\epsilon}\frac{1}{(P-q-k)^2-m_D^2+i\epsilon},
\end{eqnarray}
where $t_{D^+D^- \to D^0\bar D^0}$ is a function of the $ D^0\bar D^0$ invariant mass.
We can now take the propagators of Eq. (\ref{eq:propagator}) and perform the $q^0$ integration analytically using Cauchy's integration, and we find, keeping only the $\epsilon^i(\gamma)$ transverse components that we shall have in the Coulomb gauge ($i, j$=1, 2, 3)
\begin{eqnarray}\label{eq:amptot3}
t_L&=&-2e g_\psi t_{D^+D^- \to D^0\bar D^0} \epsilon^i (\psi) \epsilon^j (\gamma)\, 4 \int \frac{d^3q}{(2\pi)^3} \frac{1}{2\omega({\bm q})}\,\frac{1}{2\omega({\bm P}-{\bm q})} \,\frac{1}{2\omega({\bm P}-{\bm q}-{\bm k})}\nonumber\\
&\times& (q_iP_j-q_iq_j)
\frac{1}{P^0-\omega({\bm q})-\omega({\bm P}-{\bm q})+i\epsilon}
\frac{1}{P^0-\omega({\bm q})-k^0-\omega({\bm P}-{\bm q}-{\bm k})+i\epsilon}
\end{eqnarray}
and in order to get the $d$ coefficient we must look at the $k^iP^j$ component of this integral. The first term in Eq. (\ref{eq:amptot3}), with $q_i P_j$, provides this structure immediately
since when ${\bm P}\to 0$ for the rest frame of the $\psi$, the integral only depends on $k$, hence $\int d^3q \, q_i f({\bm q},{\bm k})=k_i \int d^3q \,({\bm k} \cdot {\bm q})/ {\bm k}^2 f({\bm q},{\bm k})$. The second integral for $q_i q_j$ is a bit more involved, and since we will have ${\bm P}\to 0$ at the end, we can make an expansion for ${\bm P}$ small of all the terms. Then we have an integral at the end of the type
\begin{eqnarray}
\int d^3q q_i q_j q_l P_l f(\bm{q},\bm{k})=P_l\{A'(\delta_{ij} k_l+\delta_{il}k_j+\delta_{jl}k_i)+B' k_i k_j k_l\}
\end{eqnarray}
and only the $\delta_{jl} k_i$ will contribute to the term $k_i P_j$ that we look for. Finally we find
\begin{eqnarray}
d=d_1+d_2
\end{eqnarray}
with
\begin{eqnarray}
d_1=-8 \, e\, g_\psi \, \frac{1}{{\bm k}^2} \, \int \frac{d^3 q}{(2\pi)^3} \, {\bm q}\cdot {\bm k} \,
\frac{1}{2 \omega_1} \, \frac{1}{2 \omega_1} \,\frac{1}{2 \omega_2}
\frac{1}{P^0-2\omega_1 +i\epsilon}
\frac{1}{P^0- k^0-\omega_1 - \omega_2 +i\epsilon} \, t_{D^+ D^-,D^0 \bar{D}^0} \,
\end{eqnarray}
\begin{eqnarray}
   d_2 & =& 4 \, e\, g_\psi \, \frac{1}{{\bm k}^2}\, \int \frac{d^3 q}{(2\pi)^3} \, {\bm q}\cdot {\bm k} \,
 \left\{{\bm q}^2-\frac{({\bm q}\cdot {\bm k})^2}{{\bm k}^2}  \right\}
 \,\frac{1}{2 \omega_1} \, \frac{1}{2 \omega_1} \, \frac{1}{2 \omega_2}
 \frac{1}{P^0-2\omega_1 +i\epsilon} \,  \frac{1}{P^0- k^0-\omega_1 - \omega_2 +i\epsilon} \,   \nonumber\\
&\times& \left\{ \frac{1}{ \omega_1^2} + \frac{1}{ \omega_2^2}-\frac{1}{\omega_1(P^0-2\omega_1 +i\epsilon)}
-\frac{1}{\omega_1(P^0- k^0-\omega_1 - \omega_2 +i\epsilon)} \right\} \, t_{D^+ D^-,D^0 \bar{D}^0} \,
    \end{eqnarray}
with $\omega_1=\sqrt{\bm{q^2}+m_D^2}$ and $\omega_2=\sqrt{(\bm{q}+\bm{k})^{2}+m_D^2}$.
The loop $t$ matrix for $\psi(3770) \rightarrow  \gamma D^0 \bar{D}^0$ can then be put as
\begin{eqnarray}\label{eq:dpsi}
t_{\psi(3770) \rightarrow  \gamma D^0 \bar{D}^0}=d \,M_{\psi}k \bm{\epsilon} (\psi) \cdot \bm{\epsilon} (\gamma)   \,.
\end{eqnarray}
For the case of $\psi(3770) \rightarrow \gamma D^+ D^-$ we have the same formalism for the loop substituting  $t_{D^+ D^-,D^0 \bar{D}^0}$ by
$t_{D^+ D^-,D^+ D^-}$, but we have to add the tree level contribution of Eq. (\ref{eq:treeamplitude}).

Given the structure of the amplitude it is convenient to evaluate the phase space in terms of the energy of the photon and the $D^0$ both in the $\psi$ rest frame and we obtain at the end,

\begin{eqnarray}
\frac{d \Gamma}{d M_{\rm inv}(D^0 \bar{D}^0)}= \frac{1}{8 M^2_\psi} \,
 \frac{M_{\rm inv}(D^0 \bar{D}^0)}{(2\pi)^3}  \int dE_1  \overline{\sum}\sum |t|^2 \, {\Theta(1-A^2)} \, {\Theta(M_\psi-k-E_1)}   \,,
\end{eqnarray}
where $E_1$ is the energy of the $D^0$ and $A$ is the cosine of the angle between
 the photon and $D^0$ given by
\begin{eqnarray}\label{eq:Ac}
A \equiv \cos\theta{({\bm p_1},{\bm k})}=\frac{1}{2 p_1 k}\left\{(M_\psi-k-E_1)^2-m^2_D-{\bm p^2_1}-{\bm k^2}\right\}
\end{eqnarray}
The sum and average over spins of $|t|^2$ is given for the case of $D^+ D^-$ production by
 \begin{eqnarray}\label{eq:tt1}
\overline{\sum}\sum |t|^2 &=& \frac{1}{3}(2\,e\,g_\psi)^2
\left\{
2|1+t'^A_L+t'^B_L|^2
+{\bm p^2_2}\left(\frac{1}{p_1 \cdot k} \right)^2 \left({\bm p^2_1}-\frac{({\bm p_1}\cdot {\bm k})^2}{{\bm k}^2}  \right)
+{\bm p^2_1}\left(\frac{1}{p_2 \cdot k} \right)^2 \left({\bm p^2_2}-\frac{({\bm p_2}\cdot {\bm k})^2}{{\bm k}^2}  \right)\right. \nonumber\\
 &+& \left. \left[{\bm p_1}\cdot {\bm p_2}-\frac{({\bm p_1}\cdot {\bm k})({\bm p_2}\cdot {\bm k})}{{\bm k^2}} \right]
 \left[-2\, Re(1+t'^A_L+t'^B_L) \left(\frac{1}{p_1 \cdot k}+\frac{1}{p_2 \cdot k}\right) + 2\,{\bm p_1}\cdot {\bm p_2}\frac{1}{p_1 \cdot k}\frac{1}{p_2 \cdot k} \right] \right\}
 \end{eqnarray}
where
\begin{eqnarray}\label{eq:tt2}
t'^A_L+t'^B_L \to \frac{1}{2\,e\,g_\psi} d M_{\psi} k
\end{eqnarray}

For the case of  $D^0 \bar{D}^0$ production only $t'^A_L+t'^B_L$ has to be kept and the $\overline{\sum}\sum |t|^2$ gives us $\frac{2}{3}|d M_{\psi} k|^2$, which is what we directly obtain from Eq. (\ref{eq:dpsi}).
All terms appearing in  Eq. (\ref{eq:tt1}) can be calculated in terms of    $M_{\rm inv}(D^0 \bar{D}^0)$, $E_1$ and $A$ of Eq. (\ref{eq:Ac}).

\subsection{$D\bar{D} \to D\bar{D}$ transition amplitude}

We use the Bethe-Salpeter equation in coupled channels
\begin{eqnarray}
T=[1-VG]^{-1}V  \,,
\end{eqnarray}
where the channels used are $D^+D^-$, $D^0 \bar{D}^0$,  $D_s \bar{D_s}$, plus
the $\eta\eta$ channel to account for the decay of the $D\bar{D}$
bound state into light meson-meson channels, as done in \cite{daixie}.
The $V_{ij}$ coefficients between $D^+D^-$, $D^0 \bar{D}^0$,  $D_s \bar{D_s}$
are taken from \cite{dani} and we take $V_{D^+D^-,\eta\eta}=a$,
$V_{D^0 \bar{D}^0,\eta\eta}=a$ and all the other matrix elements zero \cite{daixie}.
The value of $a$ is chosen at $a=42$ in order to obtain a width
$\Gamma \simeq 36$ MeV as found in \cite{xiaothree}. The $G$ function is
a diagonal function of the meson meson loops for which  either
dimensional regularization or cut off regularization can be used.
If the cut off regularization is used, $q_{\rm max}=830$ MeV. As discussed above,
the $q_{\rm max}$ has to be also implemented in the ${\bm q}$ integral of the loop
function. It is also interesting to note that since the $\eta\eta$ channel is only
introduced to produce the width, it is sufficient to take
\begin{eqnarray}
G_{\eta\eta}=-i\frac{1}{8\pi}\frac{1}{M_{\rm inv}} q_{\eta} \,; \qquad q_{\eta}=\frac{\lambda^{1/2}(M^2_{\rm inv},m^2_{\eta},m^2_{\eta})}{2 M_{\rm inv}}
\end{eqnarray}
The contribution of the real part of $G_{\eta\eta}$ only
introduces negligible changes in the results \cite{wangliang}.

\section{Results}

\begin{figure}[h!]
 \centering
 \includegraphics[width=11.cm]{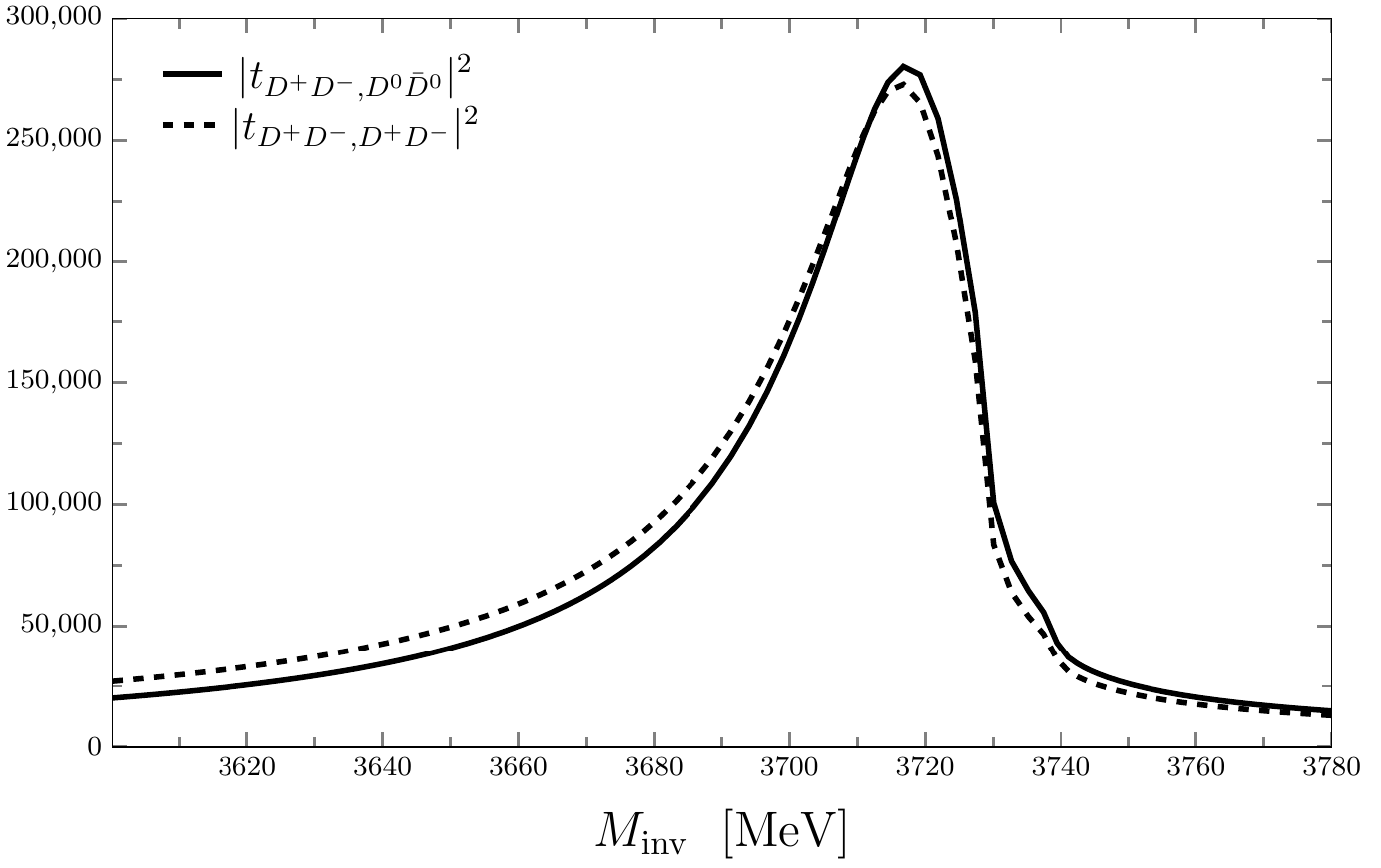}
 \caption{$|t|^2$ for $D^+ D^- \to D^0\bar{D}^0$ and $D^+ D^- \to D^+ D^-$  as a function of the $D\bar{D}$ invariant mass.}
 \label{fig:GT42}
\end{figure}

First we look at the $t_{D^+D^-,D^+D^-}$ and $t_{D^+D^-,D^0 \bar{D}^0}$ amplitudes. In Fig. \ref{fig:GT42}  we plot $|t_{D^+D^-,D^+D^-}|^2$ and $|t_{D^+D^-,D^0 \bar{D}^0}|^2$  as a function of the $D\bar{D}$ invariant mass.
We can see that the amplitudes are practically identical and have a peak around  $3770$ MeV corresponding to a $D\bar{D}$ bound state. This is due to the dominance of the isospin $I=0$ contribution.
The amplitudes also exhibit a fast  fall around threshold corresponding to the opening of the $D\bar{D}$ decay channel (Flatt\'{e} effect).

\begin{figure}[h!]
 \centering
 \includegraphics[width=11.5cm]{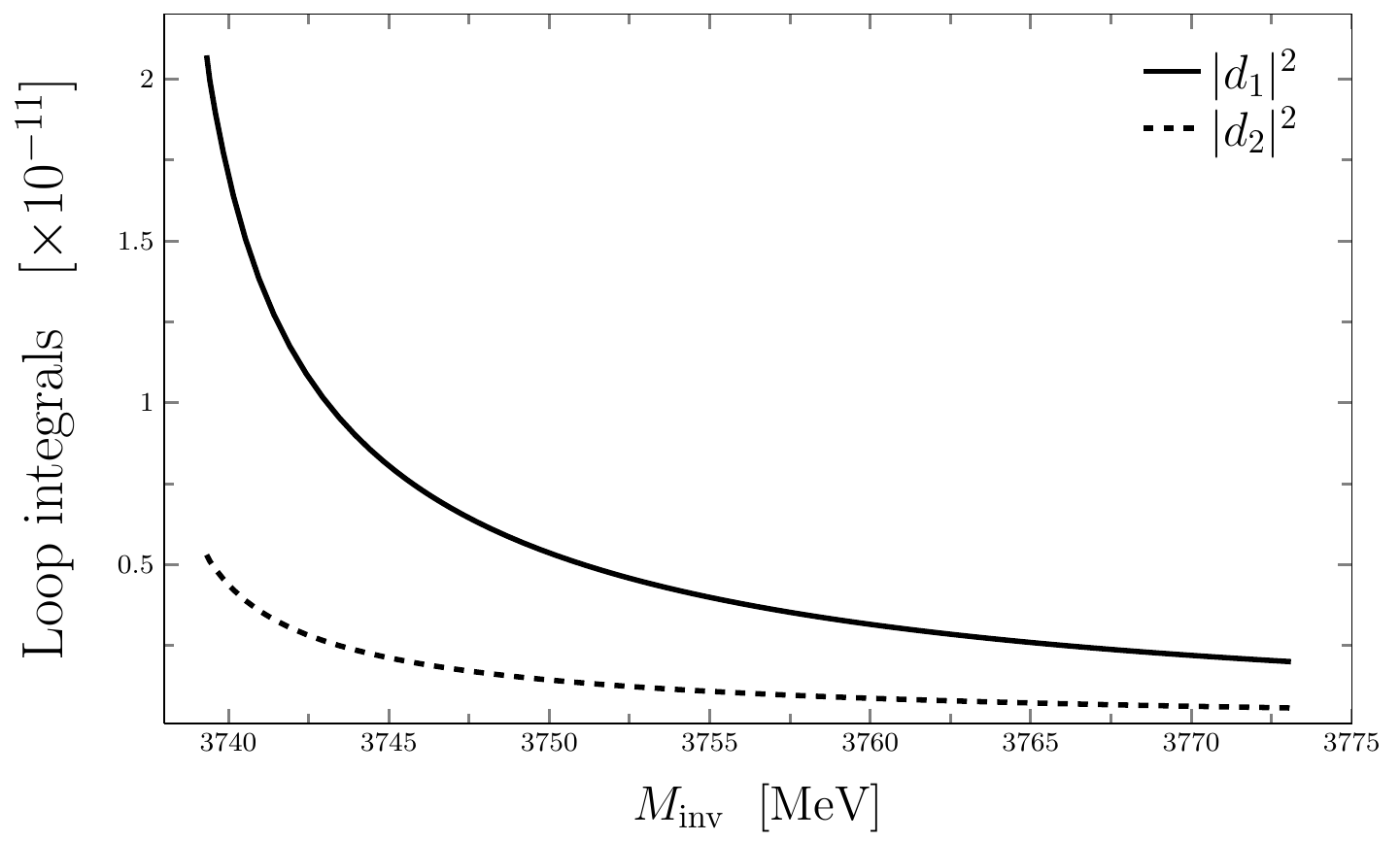}
 \caption{The modulus squared of the coefficients $d_1$, $d_2$ as a function of the $D\bar{D}$ invariant mass.}
 \label{fig:d1d2}
\end{figure}

Next we plot the coefficients $d_1$, $d_2$ in Fig. \ref{fig:d1d2} given by  Eqs. (\ref{eq:tt1}), (\ref{eq:tt2}). We see that the  coefficient $d_1$ is bigger than $d_2$  by about a factor of two. At low invariant mass close to threshold they increase, reflecting the amplitude $t_{D\bar{D}, D\bar{D}}$  which is contained in the coefficients.

\begin{figure}[h!]
 \centering
 \includegraphics[width=11.cm]{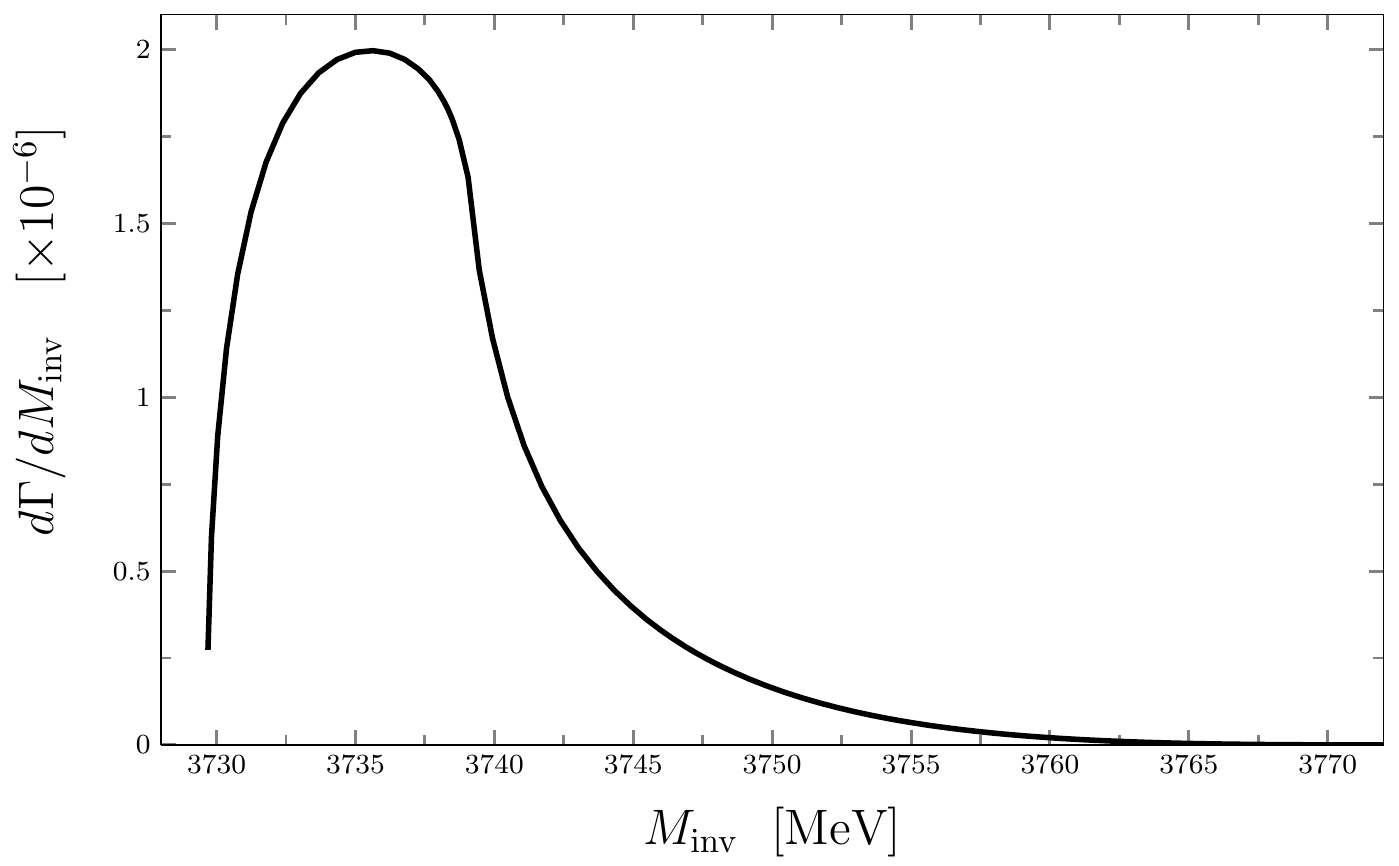}
 \caption{The differential cross section
 for $\psi(3770) \rightarrow \gamma D^0 \bar{D}^0$ as a function of the $D^0 \bar{D}^0$ invariant mass, $a=42$.
 }
 \label{fig:nudg}
\end{figure}

\begin{figure}[h!]
 \centering
 \includegraphics[width=11.cm]{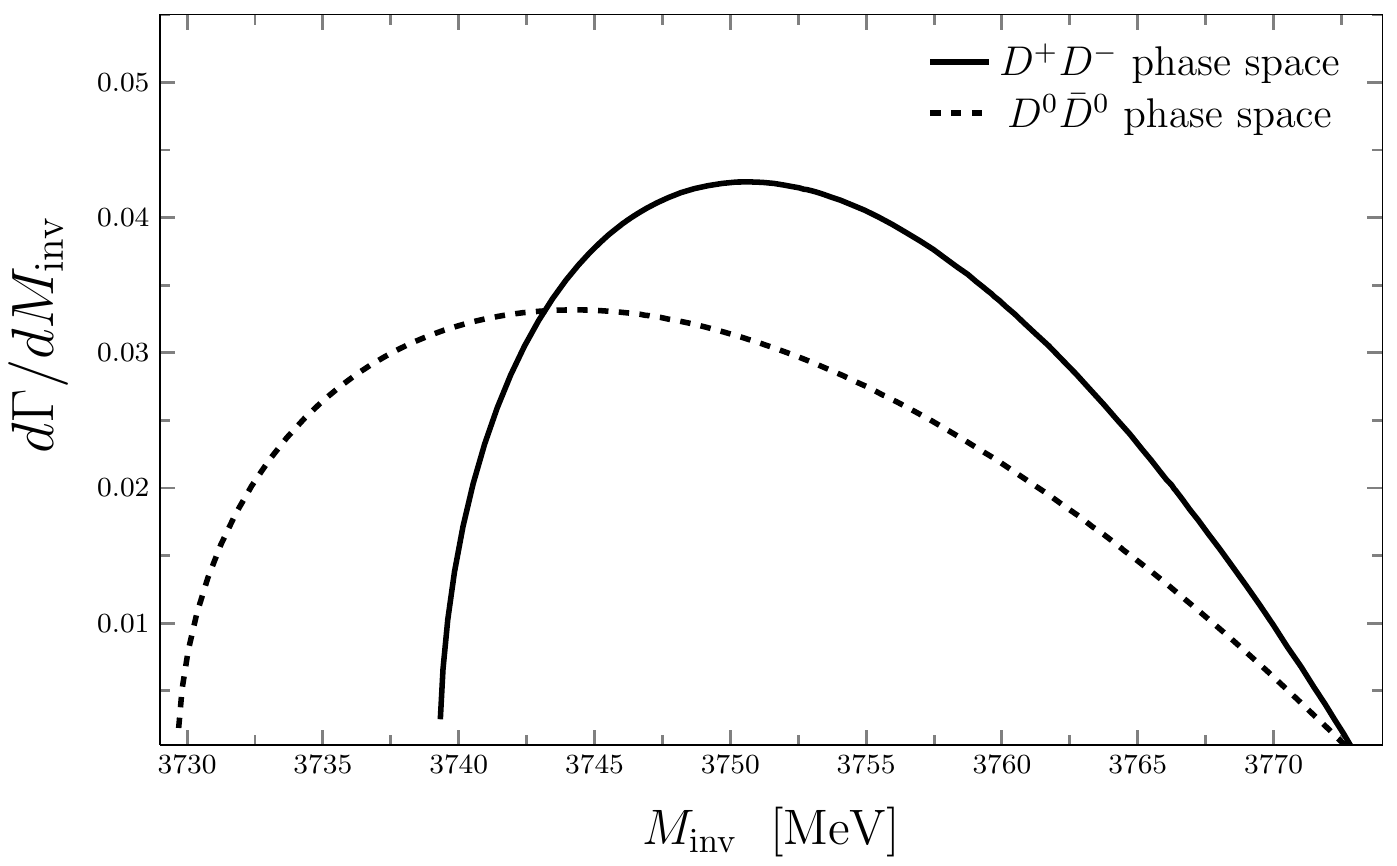}
 \caption{Phase space for $\psi(3770) \rightarrow \gamma D^0 \bar{D}^0$ and  $\gamma D^+ D^-$ normalized to the same area.}
 \label{fig:phase}
\end{figure}

The most important result is shown in Fig. \ref{fig:nudg} where we show the results of $d\Gamma/ d M_{\rm inv}$ for the
$D^0 \bar{D}^0$ distribution. We see a concentration of the strength around threshold with a peak around $3735$ MeV. In order to see that this structure is tied to the resonance below threshold we plot  in Fig. \ref{fig:phase} the phase space for $\psi(3770) \rightarrow \gamma D^0 \bar{D}^0$ substituting
$\overline{\sum}\sum |t|^2$ of Eq. (\ref{eq:Ac}) by a constant. We also show the phase space for $\psi(3770) \rightarrow \gamma D^+ D^-$ keeping the physical masses for
the $D$ mesons. We can observe that the shape of the phase space distribution is drastically different from that predicted in the presence of a $D\bar{D}$ bound state. The phase space
peaks around $3742$ MeV instead of $3735$ MeV for the distribution
 with the $D\bar{D}$ bound state. The shapes of the fall down of the two distributions are also different, the one with the $D\bar{D}$ bound state falling as a concave
curve and the phase space as a convex one.

\begin{figure}[h!]
 \centering
 \includegraphics[width=11.cm]{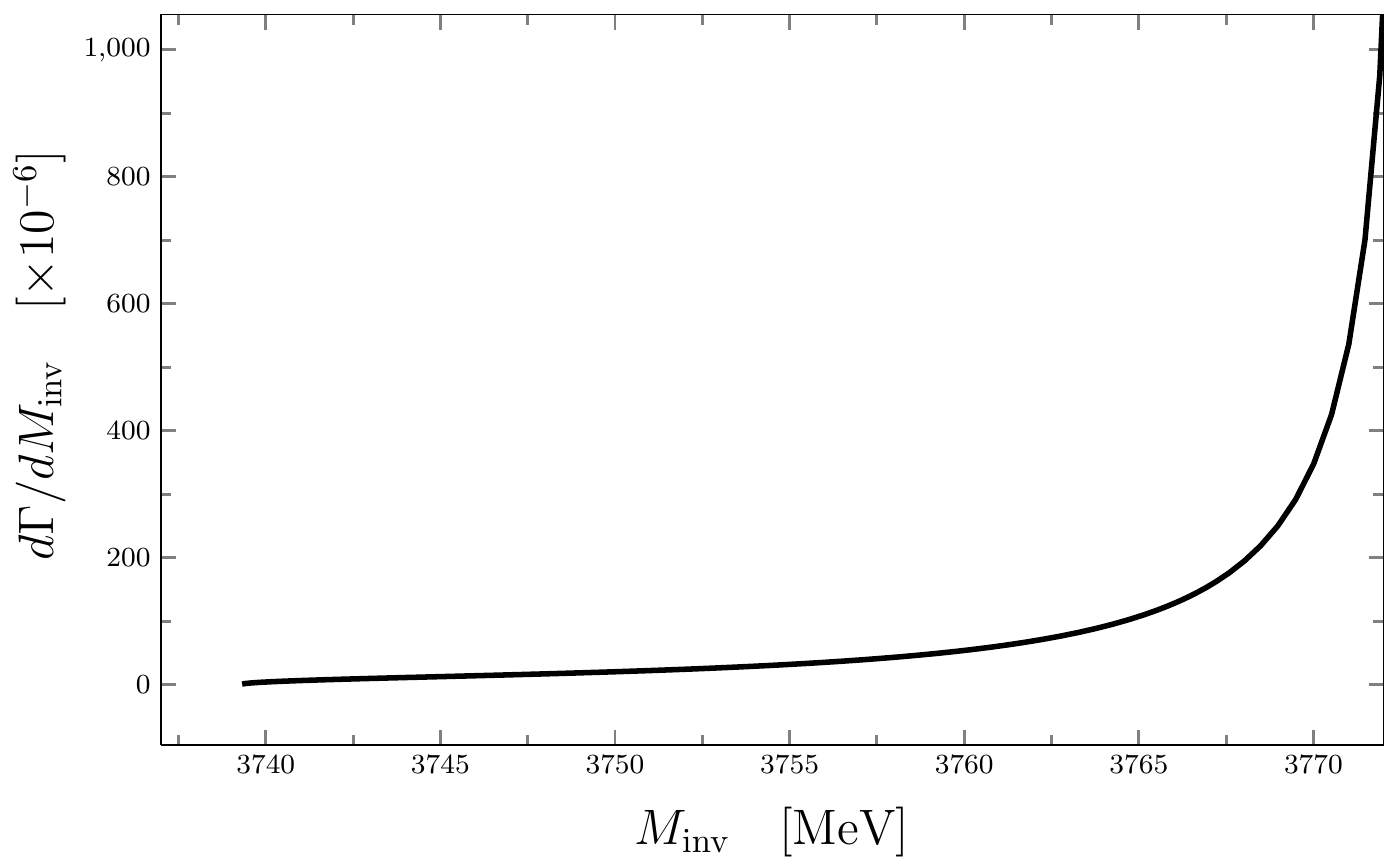}
 \caption{ The differential cross section
 for $\psi(3770) \rightarrow \gamma D^+ D^-$ as a function of the $D^+ D^-$ invariant mass.}
 \label{fig:chdg}
\end{figure}

Finally we show in Fig. \ref{fig:chdg} the mass distribution  for $\psi(3770) \rightarrow \gamma D^+ D^-$.  The shape is quite different than the one for $\psi(3770) \rightarrow \gamma D^0 \bar{D}^0$ and the reason is the contribution
of the tree level, which is absent for  $D^0 \bar{D}^0$  production. It is clear that $\psi(3770) \rightarrow \gamma D^0 \bar{D}^0$ is, thus, the reaction that better shows the presence of the $D\bar{D}$ bound state. Although not relevant for the present discussion, but we can see the
infrared divergence associated to the limit $k \to 0$
for the photon momentum, when the $D$ propagator in the tree level amplitude becomes on shell.

\newpage
The strengths of the $d\Gamma/dM_{\rm inv}$  distribution obtained fall well within present capacity at BESIII. Indeed the integrated luminosity in the $3770$ MeV  region is about
$5~ \rm fb^{-1}$ per year, 
which, together with the $e^+ e^-$ production cross section of the resonance of 7.2 nb \cite{refbo}, leads to about $3.5 \times 10^7$ accumulated
events per year.   The expected final data will have an accumulated $20~ \rm fb^{-1}$ luminosity \cite{whitepaper} and with the planned BES upgrade $15~ \rm fb^{-1}$ per year ($1.1 \times 10^8$ events per year)\footnote{Haibo Li and Changzheng Yuan, private communication.}. Our differential widths $d\Gamma/dM_{\rm inv}$ of the order of $10^{-6}$ are well measurable with these $\psi(3770)$
production rates \footnote{After completion of the work we learned that a BESIII team is presently investigating these reactions.}.


\section{Conclusions}
We have made a study of the $\psi(3770) \to \gamma D \bar D$ decay, looking at the
   $D^+ \bar D^- $ and $D^0 \bar D^0 $ mass distributions in $d\Gamma/dM_{\rm inv}$ close to theshold. We saw that the production of $D^0 \bar D^0 $ is particularly suited to study the dynamics of the $D \bar D$ interaction because the tree level contribution is zero and the process goes with a loop mechanism that involves the $D^+ D^- \to D^0 \bar D^0$ scattering amplitude. We have used the results of a theory that predicts a $D \bar D$ bound state and this has as a consequence that the $D^0 \bar D^0$ mass distribution accumulates close to threshold and diverts drastically from a phase space distribution. The rates that we obtain for the mass distribution are perfectly reachable with the present BESIII facility and we encourage the performance of the experiment that could shed light on the issue of this possible $D \bar D$ bound state, and in any case would provide information on the  $D \bar D \to D \bar D$ interaction.

\section*{Acknowledgments}
LRD acknowledges the support from the National Natural Science
Foundation of China (Grant Nos. 11975009, 11575076).
GT acknowledges the support of PASPA-DGAPA, UNAM for a sabbatical leave. This project has received funding from the European Union's Horizon 2020 research and innovation programme under grant agreement No. 824093 for the STRONG-2020 project.
This work is partly supported by the Spanish Ministerio
de Economia y Competitividad and European FEDER funds under Contracts No. FIS2017-84038-C2-1-P B
and No. FIS2017-84038-C2-2-P B.

\end{document}